\documentclass[10pt]{article}
\usepackage{Tillpack}
\title{Augmenting a Geometric Matching is $NP$-complete}
\author{Tillmann Miltzow\footnote{Institute of Computer Science, Freie Universit\"at Berlin, Germany. \texttt{t.m@fu-berlin.de}}} 

\begin{document}
\maketitle

\begin{abstract} 
	Given $2n$ points in the plane, it is well-known
	that there always exists a perfect straight-line 
	non-crossing matching.
	We show that it is $NP$-complete to decide
	if a partial matching can be augmented to 
	a perfect one, via a reduction from $1$-in-$3$-SAT. 
	This result also holds for bichromatic matchings.
\end{abstract}

\vspace{0.4cm}

	In a blitz chess tournament every player wants
	to play each round without pause and 
	play against every other player exactly once.
	To design such a tournament in mathematical terms
	means to partition the edges of the complete graph into 
	perfect matchings. 
	Matchings appear in everyday life and are
	long studied mathematical objects. 
	They give rise to 
	many natural mathematical questions and have many applications.

	
	A well known fact is that any point set in general position in the plane with an
	even number of points admits a perfect matching \cite{Kano2003B}.
	To see this go for instance by induction: Sweep a 
	vertical line $\ell$ from left to right till $2$ points are left of
	$\ell$, connect these $2$ points and continue.
	Or allow crossings and consider the matching with smallest total
	edge-length. This settles also the bicolored case.
	
	\begin{figure}[h]
			\begin{center}
				\includegraphics[width = 0.6\textwidth]{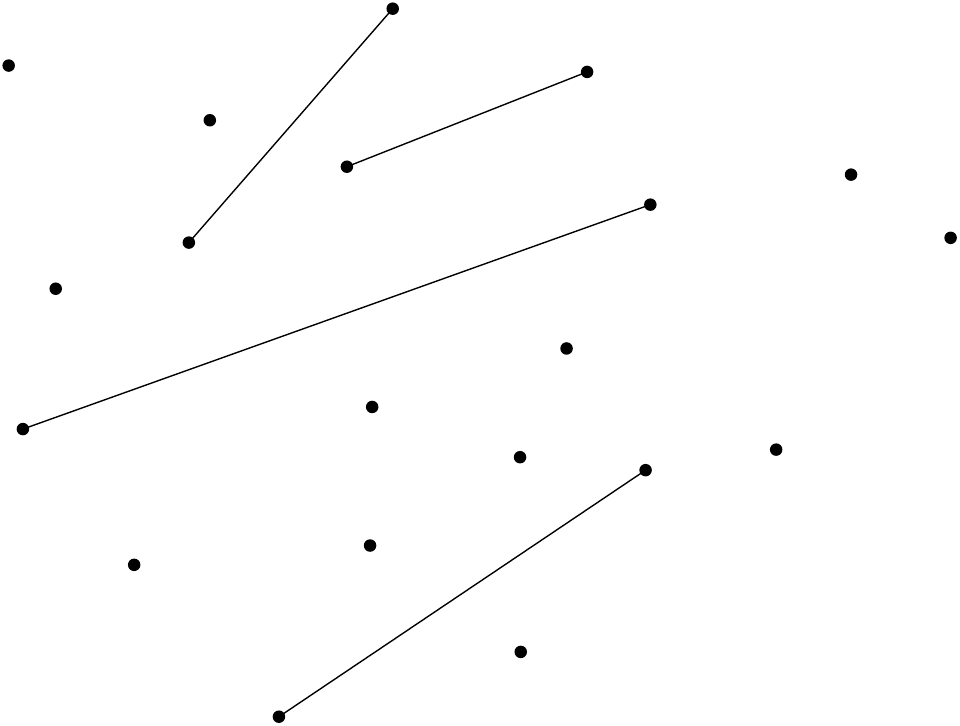}
				\caption{After some edges are drawn it is not clear,
				 how to continue. }
				\label{fig:motivation}
			\end{center}
	\end{figure}

	When one works on any kind of problem closely related to
	geometric matchings, one might take a pen and a piece of paper and
	start drawing dots representing points in the plane; just to get
	an intuition on whatever problem one is thinking about.
	One might continue to draw edges arbitrarily in a not
  too simple way, in the hope to see something or get a new idea. 
	But then one realizes
	that there are only few points left and one has
	problems to continue adding edges without introducing crossings. 
	Of course one might cheat and just draw more points at appropriate places.
	Sometimes one must do that. See Figure \ref{fig:motivation}.

	Because matchings are so natural and important mathematical objects,
	we were motivated to study the augmentability problem. 
	For related augmenting problems we refer to a recent survey from 
	Ferran Hurtado and Csaba D. T\'oth \cite{FerranToth2003survey}.
	After we present the proof of our main Theorem \ref{AugM}, we will
	present a simple application in Theorem \ref{thm:cubic}, to illustrate
	its versatility.
	Theorem \ref{thm:cubic} has already been shown by Alexander~Pilz~\cite{Augcubic2003}.
	
	
	To avoid confusions we repeat some definitions. 
	A \emph{PSLG} or \emph{planar straight line graph} consists
	of points and line segments in the plane representing vertices 
	and edges of an abstract graph. 
	A PSLG is a \emph{partial (geometric) matching} if each vertex is incident
	to at most one edge. In a \emph{perfect matching} every
	vertex is incident to exactly one edge.
	We say a PSLG is \emph{bichromatic} if every vertex has
	color \emph{red} or \emph{blue} and no two vertices
	with the same color have a common edge.
	In this paper we say that a PSLG $G=(V,E)$ augments $G'=(V',E')$
	if they have the same vertex set and  $E' \subseteq E$.
	A PSLG is \emph{cubic} if every vertex is incident to exactly $3$ edges.

	\begin{theorem}[Augmenting Matchings]\label{AugM}
		It is $NP$-complete to decide whether a partial geometric
		matching can be augmented to a perfect one, both for the colored
		and uncolored case.
	\end{theorem}
\noindent \emph{Proof:} 
	We will prove the colored case first and show later
	how the proof carries over to the uncolored case.
	We will reduce $1$-in-$3$-SAT to the decision problem above.
	It can be checked in linear time whether 
	a given set of edges augments a matching. This is true for the colored and uncolored case.
	Thus the decision problem lies in $NP$.
	We will first describe gadgets and then 
	describe how to encode an instance of $1$-in-$3$-SAT in an instance
	of our augmentability problem using these gadgets \cite{DBLP:books/daglib/0086373}.

		\begin{figure}[h]
			\begin{center}
				\includegraphics[width = \textwidth]{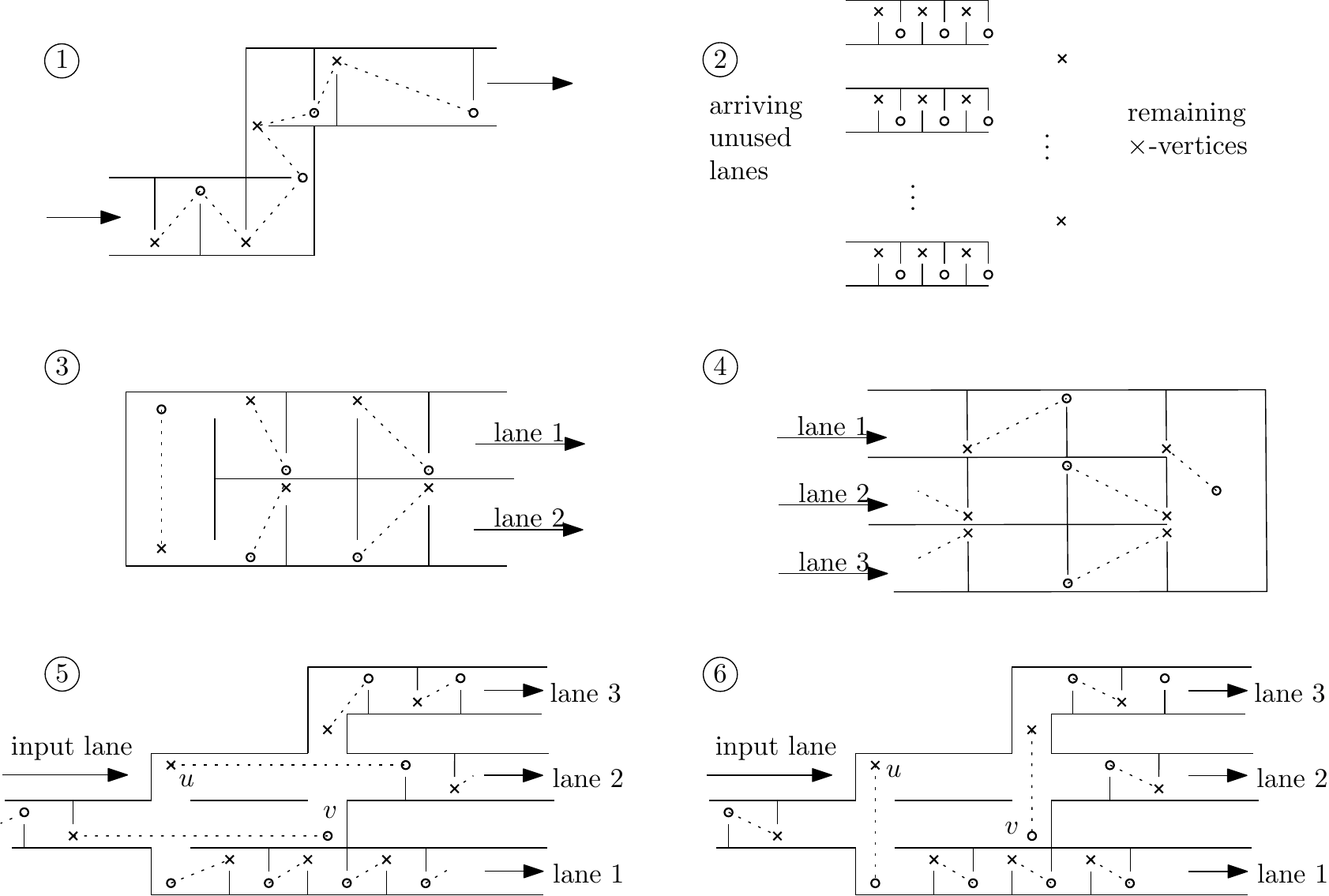}
				\caption{
				1: The lane surrounds the free vertices 
					such that each vertex has only two possible neighbors
					it can connect to; 
				2: The basin consists of all the
					lanes that we did not use.
				3: The variable gadget consists of $2$ free vertices of
					different color, each emits one lane;
				4: The clause gadget consists of exactly one $\medcirc$-vertex
					which can be matched to exactly one of the arriving lanes;
				5\&6: The multiplier has one lane
				arriving from the left and three lanes emit.
				Depending on the information, the lane carries, there is
				only one way to match the edges. In each case
				lane $1$ carries the opposite information as 
				the input lane and lanes $2$ and $3$ carry the same
				information as the input lane.
					 }
				\label{fig:firstGadgets}
			\end{center}
	\end{figure}
	
\begin{figure}[ht]
			\begin{center}
				\includegraphics[width = \textwidth]{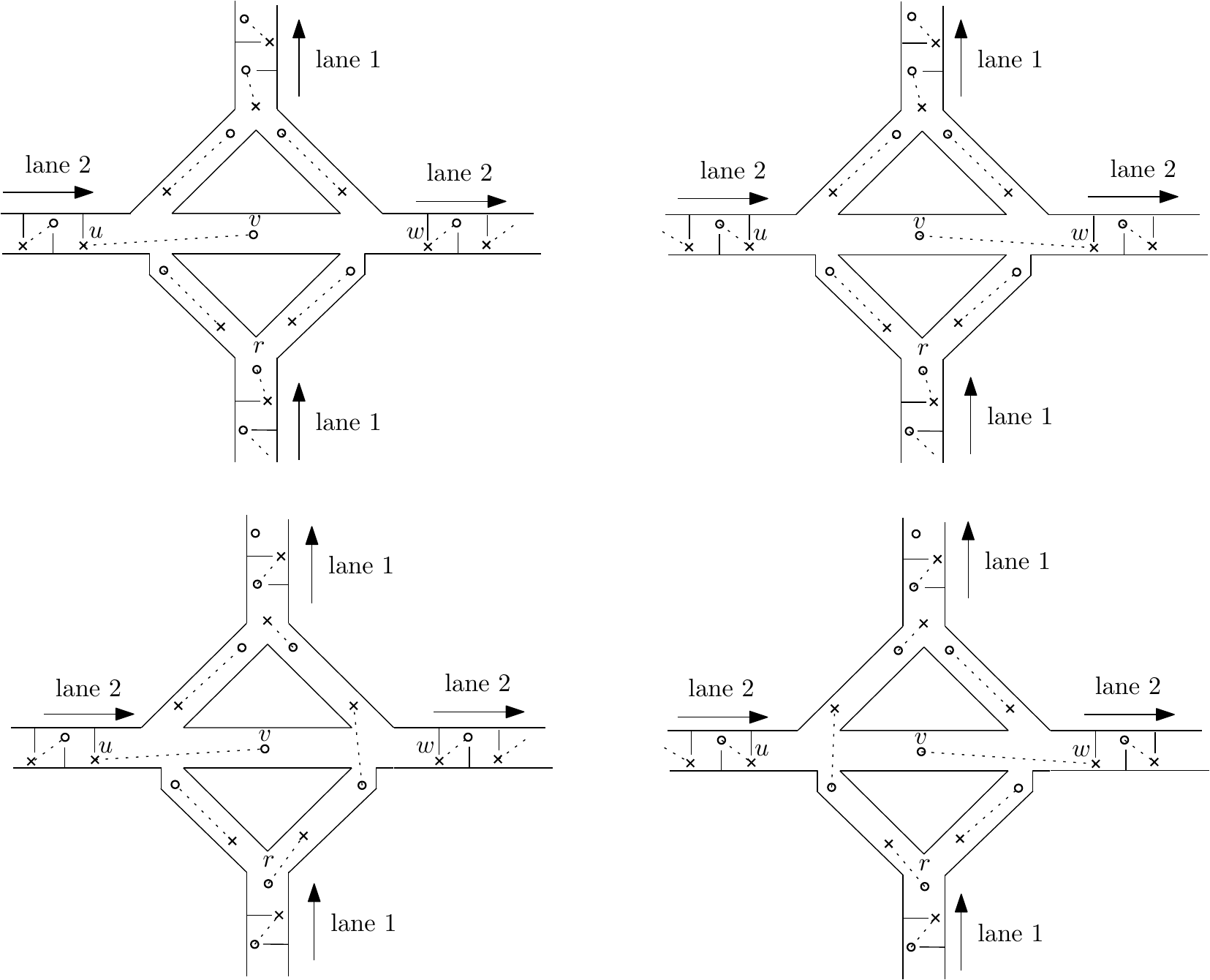}
				\caption{The junction is needed to let two lanes cross. There
				are four possible combinations of information the lanes carry.
				For each combination there is only one way 
				to match the free vertices and the information is always carried over to
				the other side of the junction.}
				\label{fig:junction}
			\end{center}
	\end{figure}

	The construction not surprisingly consists of 
	\emph{variable gadget}s and \emph{clause gadget}s. We will use
	\emph{lane}s to transport the information from
	the variable gadgets to the clause gadgets.
	Lanes might have to cross each other on their way\footnote{Even when we considered a reduction from \emph{planar} $1$-in-$3$-SAT \cite{DBLP:books/daglib/0086373}, junctions were still necessary.}
	from the variable gadget to the clause gadget and at other points which will become apparent later.
	For this purpose \emph{junction gadget}s 
	or just \emph{junction}s will be introduced. 
	To encode the negation of a variable, we have	certain \emph{multiplier-gadget}s or
	just \emph{multiplier}s, 
	which also serve for us to split a lane in two.
	This is necessary as  our variable gadgets only emit two lanes
	initially. Thus, we can create as many lanes as we want
	with the multiplier gadget to transport the information
	from a certain variable. 
	This is necessary if a variable appears in more than one clause
	of the original formula.
	In the end we will have the problem that some lanes will not be used. 
	We let all of them go into a common \emph{basin},
	where they can be taken care of easily.
	See Figure \ref{fig:firstGadgets} and Figure \ref{fig:junction}
	for illustrations.
	All of our gadgets consist of the input edges
	and free vertices arranged in the plane.
	For the clarity of the drawings
	we did not draw the edges separately, but let
	them appear as one PSLG. This is not problematic 
	as one could perturb each edge slightly such that
	they do not overlap nor cross and still all the crucial properties remain.
%

	We start to describe the lane, which transports
	information. 
	The lane consists of a tube bounded
	by matching edges and unused points.
	The tube is piked with matching edges to guarantee that
  each free vertex has exactly two possible neighbors
  he could be connected to. 
	Thus any matching edge determines the matching
	of all edges in the lane and there are exactly
	two ways to match all the free vertices. 
	Also we attach to a lane a direction as we say it starts
	at a variable gadget and ends in a clause gadget or
	the basin; maybe bypassing several other gadgets in between.
	
	We say that a lane transports the information \texttt{true} if the orientation
	of the edge is $\times \medcirc$ (with respect to the direction of the lane)
	and false if it is $ \medcirc \times$.

	Note that lanes do not need to be straight, but can have any kind of bends.


	The variable gadget consists of $2$ free-vertices at the start of $2$ lanes.
	We say that	a variable is set to \texttt{true} if the two free vertices
	are matched to one another, and \texttt{false} otherwise.
	The upper lane transports the information $x$ and the lower one transports
	the information $\neg x$ by our convention.


	The basin is in fact not a real
	gadget as we only gather all the lanes to a common
	location and place sufficiently many $\times$-vertices next to these lanes to serve them
	(i.e. connect the $\medcirc$-vertices, that are not connected yet.). It holds
	$ \# (\text{$\times$-vertices in the basin}) = 
	\# (\text{$\medcirc$-vertices}) - \#(\text{$\times$-vertices}).$
	This is obvious as the total number of $\times$-vertices and $\medcirc$-vertices
	must agree.

	The clause gadget consists of exactly one free $\medcirc$-vertex
	which is exposed to three arriving lanes. 
	The $\medcirc$-vertex can only
	be matched to exactly one of the three lanes.
	Thus all free vertices can be matched iff exactly one of the lanes
	transports the information \texttt{true}.

	The multiplier gadget accepts one arriving lane and emits $3$ leaving ones.
	Lane $1$ and $2$ as shown in Figure \ref{fig:firstGadgets}
	carry the same information as the arriving lane 
	and lane $3$ will transport the opposite information.
	Observe that the vertex denoted $v$ in Figure \ref{fig:firstGadgets} can only be adjacent to two
	vertices. To which is determined by the information
	of the input lane. This in turn determines what $u$ 
	is connected to and will decide the edges for 
	the three emiting lanes.
	
	\begin{figure}[ht]
			\begin{center}
				\includegraphics[width = \textwidth]{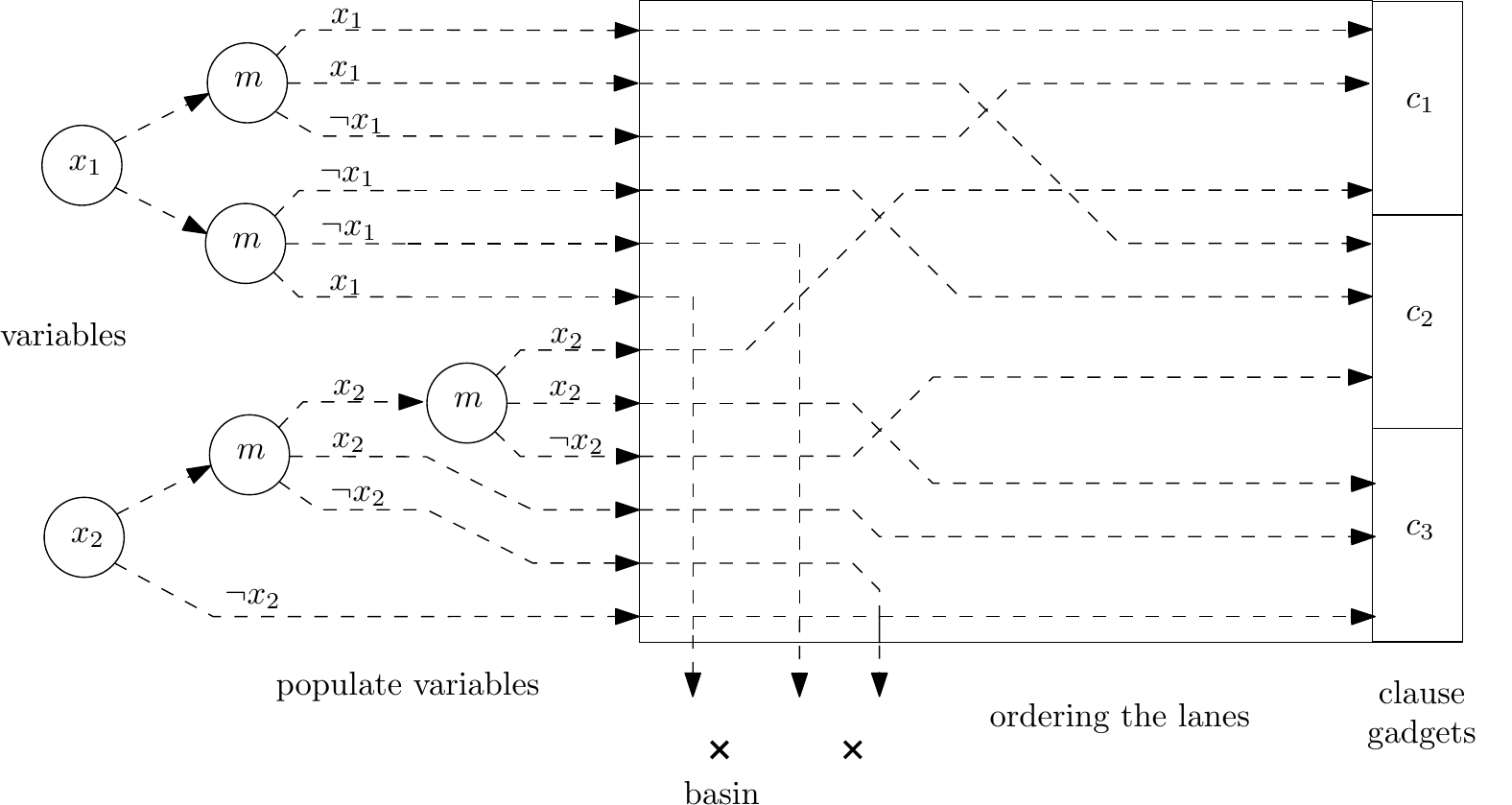}
				\caption{The construction: lanes are depicted by dashed lines; the multipliers
				are represented by a circle with an $m$; variable gadgets are represented
				by the corresponding variable names surrounded by a circle; 
				when two lanes cross a junction is implied}
				\label{fig:construction}
			\end{center}
	\end{figure}

	The junction is depicted in Figure \ref{fig:junction} together with all possible 
	assignments of the incoming lanes and a notation for the vertices. Depending on lane $2$ the point $v$ must 
	be matched with $u$ or $w$. This implies that either the left tunnel or the right tunnel is blocked
	by an edge. This is no problem when lane $1$ carries the \texttt{true} information,
	as in the top of Figure \ref{fig:junction}. But it is still no problem
	when lane $1$ carries the information
	\texttt{false}. At vertex $r$ has to be made a choice such that the
	information can be passed through the correct tunnel, as in the bottom of Figure \ref{fig:junction}. 



	We are now ready to present the whole construction, see Figure \ref{fig:construction} for an
	illustration. 
	The construction consist of 5 parts. The first part consists merely of 
	variable gadgets. The second part
	populates the lanes from the variables using the multiplier. 
	This is done till  each literal is at least as many  times present as 
	it is used in the clauses. The third part is the most important one, 
	here the literals are connected to the corresponding clauses. Note 
	that we have at most a quadratic number of crossings.
	At the bottom of the third part is the fourth part, 
	the basin, which takes care of overmuch lanes.
	The last part are the clause gadgets, where the lanes end. 
	It is clear, that this construction can be made in polynomial time. 
	Now let $\Phi$ be an instance of $1$-in-$3$-SAT and $M(\Phi)$ the construction presented. If there
	is an assignment to the variables such that $\Phi$ is \texttt{true} we can match the
	variable gadget accordingly and transport these information over the lanes
	and each clause gadget will be satisfied. Thus it is possible to augment the matching
	of $M(\Phi)$ to a perfect one. On the other hand if $M(\Phi)$ can be augmented to
	a perfect matching we get an assignment from the variable gadget, which satisfies
	each clause. 
	
	Note that we have not used the color restriction in the proof. 
	This implies that each gadget would still work if the colors 
	were removed and thus it is also $NP$-hard to decide whether a 
	partial matching of an uncolored point set can be augmented 
	to a perfect one. This completes the proof. \hspace{\stretch{1}}~\smiley

	\vspace{0.2cm}

	\begin{figure}[h]
			\begin{center}
				\includegraphics[width = 0.5\textwidth]{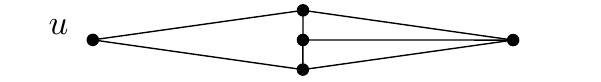}
				\caption{All vertices of this graph except $u$ have 
				$3$ incident edges}
				\label{fig:cubicGadget}
			\end{center}
	\end{figure}
	
	As promised we present now a new proof of a recent result from Alexander Pilz using Theorem \ref{AugM}.
	
	\begin{theorem}[Augmenting cubic PSLG, Pilz \cite{Augcubic2003}]\label{thm:cubic}
		It is $NP$-complete to decide whether a PSLG
		can be augmented to a cubic one.				 
	\end{theorem}
	
	\noindent \emph{Proof:}
		Clearly the problem lies in $NP$. 
		To see that it is $NP$-hard we make a reduction from the
		previous problem. Let $M$ be a partial matching.
		Replace every vertex $v$ by a small rotated copy of the gadget in Figure \ref{fig:cubicGadget}.
		The edges of $M$ remain where	they were.
		We denote the resulting PSLG by $G(M)$.
		
		It is clear that this can be done in polynomial time and
		we observe that $M$ can be augmented to a perfect matching
		iff $G(M)$ can be augmented to a cubic PSLG. \hspace{\stretch{1}}~\smiley

\section*{Acknowledgments}
	The author thanks Andrei Asinowski 
	for his help for improving the presentation.

\bibliography{TillLib}
\bibliographystyle{plain}
	
\end{document}